\documentclass[aps,prl,twocolumn,showpacs]{revtex4}
\usepackage{amsfonts,amsmath,amssymb,graphicx}
\usepackage{amsfonts}
\usepackage{amsmath}
\usepackage{amssymb}
\usepackage{graphicx}

\def\kbar{\protect\@kbar}
\def\@kbar{\relax \bgroup
\def\@tempa{\hbox{\raise.73\ht0
\hbox to0pt{\kern.25\wd0\vrule width.5\wd0 height.1pt
depth.1pt\hss}\box0}}\mathchoice{\setbox0\hbox{$\displaystyle
k$}\@tempa}{\setbox0\hbox{$\textstyle
k$}\@tempa}{\setbox0\hbox{$\scriptstyle
k$}\@tempa}{\setbox0\hbox{$\scriptscriptstyle k$}\@tempa}\egroup}

\begin{document}

\title{\textbf{Staggered-Ladder Quasienergy Spectra for Generic Quasimomentum and Quantum-Dynamical Manifestations}}
\author{Itzhack Dana}
\affiliation{Minerva Center and Department of Physics, Bar-Ilan University, Ramat-Gan
52900, Israel}

\begin{abstract}
A new kind of regular quasienergy (Floquet) spectrum is found for the generalized kicked particle under quantum-resonance conditions at generic quasimomentum, a quantity most relevant in atom-optics experimental realizations of kicked-rotor systems. The new non-Poisson regular spectrum has the structure of a staggered ladder, i.e., it is the superposition of a finite number of ladder subspectra all having the same spacing, which is independent of the nonintegrability of the system. This spectral structure is shown to have distinct quantum-dynamical manifestations: A suppression of quantum resonances and a novel type of dynamical localization characterized by unique features such as traveling-wave components in the time evolution. These phenomena are found to be robust under small variations of the quasimomentum and should therefore be experimentally observable using Bose-Einstein condensates with sufficiently small quasimomentum width. 
\newline
\end{abstract}
\maketitle

The nature of the energy and quasienergy spectra of quantum systems whose
classical limit is nonintegrable has been the subject of an enormous number
of studies during the last four decades. Classically integrable systems generally feature \textquotedblleft regular\textquotedblright\ energy spectra \cite{ip} with a Poisson level-spacing distribution \cite{bt}. For completely chaotic systems, this distribution is of Wigner type \cite{wd}. The distribution for systems with a mixed phase space is, under some assumptions, a weighted superposition of Poisson and Wigner distributions associated with the regular and chaotic phase-space regions \cite{br}.

For time-periodic quantum systems, the energy is replaced by the quasienergy
(QE), giving the eigenvalues of the one-period evolution (Floquet) operator.
Paradigmatic and realistic models are the kicked-rotor systems \cite{dl,al,ali,cs,ff,fmi,fm,fmir,qam,kp,qr,e1,e10,e2,e3,e4,e5,e6} exhibiting a variety of phenomena, the most well-known one being dynamical localization \cite{dl}, i.e., a quantum suppression of the classical chaotic diffusion for irrational values of a scaled Planck constant $\hbar _{\mathrm{s}}$. This phenomenon can be attributed to an Anderson-like localization of QE eigenstates in angular-momentum space 
\cite{al}. This localization was numerically found to imply a Poisson QE level-spacing distribution \cite{ff}. Other distributions were found for rational values of $\hbar _{\mathrm{s}}$ \cite{fmi,fm}. The QE levels for rational $\hbar _{\mathrm{s}}$ actually correspond to bands \cite{fmir}, giving a continuous QE spectrum. This leads to another well-known phenomenon, quantum resonance \cite{dl,cs,fmir}, a quadratic growth in time of the mean kinetic energy.

During the last two decades, kicked-rotor systems have been
experimentally realized using atom-optics techniques with cold atoms or
Bose-Einstein condensates (BECs) \cite{e1,e10,e2,e3,e4,e5,e6}. This allowed to
observe in the laboratory several quantum-chaos phenomena, including
dynamical localization and quantum resonance \cite{e1}, and to verify theoretical predictions. In the experiments, the kicked rotor and variants of it were actually realized as kicked-\emph{particle} systems, since atoms move on lines and not on
circles like rotors. These realizations are based on the fact that a kicked particle reduces to a generalized kicked rotor at any fixed value of the conserved \emph{quasimomentum} $\beta $ of the system \cite{kp} (see also below). The usual kicked rotor, whose QE spectral statistics has been studied as mentioned above,
corresponds to the particular case of $\beta =0$. However, several important
phenomena arise for arbitrary $\beta$ and have been experimentally
realized \cite{e2,e3,e4,e5}. Quantum resonance strictly occurs for rational values of both $\hbar _{\mathrm{s}}$ and $\beta$ \cite{kp,qr} but it has experimentally observable effects also for rational $\hbar _{\mathrm{s}}$ and general $\beta$ near the strict rational value if BECs with sufficiently small quasimomentum width are used \cite{e2,e3}.  

In this paper, we show that QE spectra for rational $\hbar _{\mathrm{s}}$ and generic $\beta $ exhibit a new kind of regularity having several quantum-dynamical manifestations which should be experimentally observable. The new \emph{non}-Poisson regularity turns out to affect significantly the QE spectra for generic $\beta$ relative to those for $\beta =0$. We consider the generalized quantum kicked particle, described by the Hamiltonian 
\begin{equation}
\hat{H}=\frac{\hat{p}^{2}}{2}+kV(\hat{x})\sum_{s=-\infty }^{\infty }\delta
(t-s),  \label{KP}
\end{equation}
where $\hat{x}$ and $\hat{p}$ are scaled dimensionless position and momentum operators ($[\hat{x},\hat{p}]=i\hbar $, with $\hbar$ denoting a scaled Planck constant), $k$ is a dimensionless nonintegrability parameter, and $V(\hat{x})$ is a general $2\pi$-periodic potential. We show that the QE spectrum of (\ref{KP}) for rational $\hbar _{\mathrm{s}}=\hbar /(2\pi )$ and generic $\beta$ has a \emph{staggered-ladder} structure, i.e., it is the superposition of a finite number of ladder subspectra all having the same spacing, which is independent of the nonintegrability $kV(\hat{x})$. For irrational $\beta$, each subspectrum is a dense ladder of levels covering all the QE range. For strict quantum resonance, i.e., rational $\beta$, each subspectrum is a finite ladder of bands and essentially \emph{no} ladder regularity occurs for $\beta =0$, see Fig. 1. Staggered-ladder spectra were first discovered for the Fokker-Planck equation \cite{sl} and consist of two ladders. In our case, the number of QE ladders can be arbitrary. We show that this QE spectral structure has distinct quantum-dynamical manifestations: A suppression of quantum resonances and a novel type of dynamical localization, basically different from the known one for irrational $\hbar _{\mathrm{s}}$ and $\beta =0$ \cite{al,dl} in several features, such as traveling-wave components in the time evolution. These phenomena are shown to be robust under small variations of $\beta$ and should therefore be experimentally observable using BECs with sufficiently small quasimomentum width.
 
\begin{figure}[tbp]
\includegraphics[width=7.5cm]{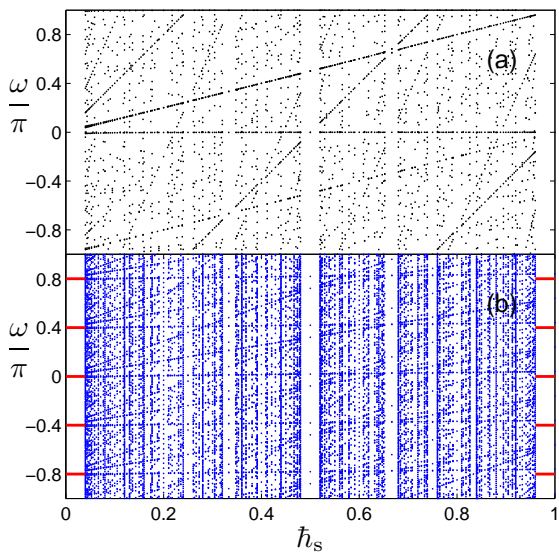}
\caption{(Color online) Dimensionless QE spectra $\omega$ of (\ref{KP}) as functions of the scaled Planck constant $\hbar _{\mathrm{s}}=\hbar /(2\pi )$ for $k/\hbar =0.1$, $V(x)=\cos (x)$, and: (a) $\beta =0$ (usual kicked rotor); (b) $\beta =0.2$, featuring spectral ladders with dominant spacing $\Delta \omega =2\pi /5$. These ladders are indicated by the left and right solid (red) segments. In both cases, $\hbar _{\mathrm{s}}$ takes all rational values in $[0,1)$ with denominators $\leq 25$.}
\label{fig1}
\end{figure}
   
We first summarize relevant known facts \cite{kp} about the system (\ref{KP}). The one-period evolution operator for (\ref{KP}) is
\begin{equation}
\hat{U}=\exp [-i\hat{p}^{2}/(2\hbar )]\exp [-ikV(\hat{x})/\hbar ].  \label{U}
\end{equation}
The QE states $\Psi _{\omega }(x)$, with the QE $\omega $ ranging in the interval $0\leq \omega <2\pi $, are the eigenstates of (\ref{U}):
\begin{equation}
\hat{U}\Psi _{\omega }(x)=\exp (-i\omega )\Psi _{\omega }(x).  \label{qee}
\end{equation}
The $2\pi $-periodicity of (\ref{U}) in $\hat{x}$ implies that $\Psi
_{\omega }(x)$ can be chosen to have the Bloch form: 
\begin{equation}
\Psi _{\omega }(x)=\exp (i\beta x)\psi _{\beta ,\omega }(x),  \label{bf}
\end{equation}
where $\beta $ is the quasimomentum ($0\leq \beta <1$), whose meaning is briefly explained below, and $\psi_{\beta ,\omega }(x)$ is $2\pi $-periodic in $x$. After inserting (\ref{bf}) into Eq. (\ref{qee}), one easily finds \cite{kp} that $\psi _{\beta ,\omega }(x)$ is an eigenstate of
\begin{equation}
\hat{U}_{\beta}=\exp [-i(\hat{p}+\beta\hbar )^{2}/(2\hbar )]\exp [-ikV(\hat{x})/\hbar ] \label{ub}
\end{equation}
with eigenvalue $\exp(-i\omega)$. Due to the $2\pi$-periodicity of $\psi_{\beta ,\omega }(x)$, one can interpret $\hat{p}$ in Eq. (\ref{ub}) as an angular-momentum operator with eigenvalues $n\hbar$ ($n$ integer). Then, $\hat{U}_{\beta}$ is the evolution operator of a \textquotedblleft $\beta$-kicked-rotor\textquotedblright\ with angle $\theta =x$ and $\beta$ is conserved during the evolution. To illustrate this conservation and the physical meaning of $\beta$, assume an initial momentum state $\phi (x) =\exp (ipx/\hbar )$. This can be written in the form (\ref{bf}) as $\exp (i\beta x)\exp (inx)$, where $\beta$ and $n$ are, respectively, the fractional and integer parts of $p/\hbar$. Then, after $s$ kicks, this state will evolve to a state having still the form (\ref{bf}) with the same $\beta$: $\phi (x;s)=\exp (i\beta x)\varphi_{\beta} (x;s)$, where $\varphi_{\beta} (x;s)$ is $2\pi $-periodic in $x$. The usual kicked rotor corresponds to $\beta =0$.

We now introduce the operators $\hat{T}_{j}=\exp (2\pi ij\hat{x}/\hbar )$ for
all integers $j$. From $[\hat{x},\hat{p}]=i\hbar $ or $\hat{x}=i\hbar d/dp$, we see that $\hat{T}_{j}$ is just a translation $\exp (-2\pi jd/dp)$ in momentum by $-2\pi j$. This implies that 
\begin{equation}
\hat{T}_{j}e^{-i\hat{p}^{2}/(2\hbar )}=e^{-i\hat{p}^{2}/(2\hbar )}e^{-2i\pi ^2
j^{2}/\hbar }e^{2\pi ij\hat{p}/\hbar }\hat{T}_{j}.
\label{iden}
\end{equation}
In turn, since $\hat{p}=-i\hbar d/dx$, $\exp (2\pi ij\hat{p}/\hbar )$ in Eq. (\ref{iden}) is a translation $\exp (2\pi jd/dx)$ in $x$ by $2\pi j$. Like $\hat{T}_{j}$, this translation obviously commutes with $V(\hat{x})$. Then, by applying $\hat{T}_{j}$ to both sides of (\ref{qee}), using (\ref{bf}), (\ref{iden}), and the notation $\hbar _{\mathrm{s}}=\hbar /(2\pi )$, we easily find that the state
\begin{equation}
\hat{T}_{j}\Psi _{\omega }(x) = \exp[i(\beta +j/\hbar _{\mathrm{s}})x]
\psi _{\beta ,\omega}(x) \label{tqe} 
\end{equation}
is an eigenstate of $\hat{U}$ with QE 
\begin{equation}
\omega _{j}=\omega +2\pi j\beta +\pi j^{2}/\hbar _{\mathrm{s}} \ \ 
\mathrm{mod}(2\pi ).  \label{wj}
\end{equation}
In addition, the state (\ref{tqe}) can be written as a Bloch state $\exp (i\beta_j x)\psi^{(j)}_{\beta_j ,\omega }(x)$ having quasimomentum
\begin{equation}
\beta _{j}=\beta + j/\hbar _{\mathrm{s}} \ \ \mathrm{mod}(1)  \label{bj}
\end{equation}
and a $2\pi$-periodic part (eigenstate of the $\beta_j$-kicked-rotor)   
\begin{equation}
\psi^{(j)}_{\beta_j ,\omega }(x)= \exp (in_j x)\psi_{\beta ,\omega }(x),
\label{ppj}
\end{equation}
where $n_j$ is the integer part of $\beta +j/\hbar _{\mathrm{s}}$.

For rational $\hbar _{\mathrm{s}}=l/q$ ($l$ and $q$ are coprime integers) and $j=rl$ ($r$ arbitrary integer), Eqs. (\ref{wj})-(\ref{ppj}) reduce to
\begin{equation}
\omega _{rl}=\omega +2\pi rl(\beta +q/2)\ \mathrm{mod}(2\pi ),\ \ \beta_{rl}=\beta ,  \label{wr}
\end{equation}
\begin{equation}
\psi^{(rl)}_{\beta,\omega }(x)= \exp (irqx)\psi_{\beta ,\omega }(x). 
\label{pprl}
\end{equation}
Eq. (\ref{wr}) means that the levels $\omega_{rl}$ form a QE ladder subspectrum of the $\beta$-kicked-rotor, with spacing independent of the nonintegrability. For irrational $\beta$, the ladder (\ref{wr}) covers densely the entire QE range. We show in what follows that $\omega$ in Eq. (\ref{wr}) can take only $q$ independent values, so that the full QE spectrum of the $\beta$-kicked-rotor is the superposition of $q$ ladders (\ref{wr}). We start from the case of rational $\beta $. In this case, we see that a ladder (\ref{wr}) consists of $g$ levels with spacing $\Delta \omega =2\pi /g$, where $g$ is the smallest integer such that $gl(\beta +q/2)$ is integer; see, e.g., Fig. 1(b) for $\beta =1/5$. For $\beta =0$, there are either no ladders ($g=1$ for $q$ even) or trivial ladders with spacing $\Delta\omega =\pi$ ($g=2$ for $q$ odd), so that almost no ladder regularity occurs, see Fig. 1(a). 

Next, it is easy to check that the momentum translation $\hat{T}_{gl}=\exp (igq\hat{x})$ commutes with the $\beta$-kicked-rotor evolution operator (\ref{ub}). Then, using standard methods \cite{cs,fmir,id}, one can find simultaneous eigenstates of $\exp (igq\hat{x})$ and $\hat{U}_{\beta}$ with respective eigenvalues $\exp (igq\alpha)$ and $\exp [-i\omega_b(\alpha ,\beta )]$. Here $0\leq\alpha <2\pi /(gq)$ and $b=1,...,gq$ is an index labeling $gq$ QE bands $\omega_b(\alpha ,\beta )$, each spanned by $\alpha$. These $gq$ bands give the full QE spectrum of $\hat{U}_{\beta}$. Since any QE $\omega$ must belong to a ladder subspectrum (\ref{wr}) of $g$ levels, it follows that the QE spectrum is the superposition of $q$ ladders of $g$ bands each; such a ladder is given by Eq. (\ref{wr}) with $\omega$ being one of $q$ bands $\omega_{c}(\alpha, \beta )$, $c=1,...,q$. As $gq\rightarrow\infty$, the width of each band should decrease and vanish as for the usual kicked rotor ($\beta =0$, $g=1$ or $2$) \cite{fmir}. This holds exactly for $q=1$ and general potential $V(\hat{x})$ \cite{qr}. We have extensively checked numerically that it holds also for $q>1$. The $q$ finite band ladders will then reduce to $q$ dense ladders of levels in the limit $g\rightarrow\infty$ of irrational $\beta$. 

Let us now consider quantum-dynamical manifestations of this staggered-ladder QE spectrum. A band continuous spectrum leads to quantum resonance \cite{fmir,kp,qr,e2}, a quadratic growth in time of the mean kinetic energy of the $\beta$-kicked-rotor. Due to the decrease and vanishing of the bandwidth as $g\rightarrow\infty$, the $l/q$ quantum resonance for rational $\beta\neq 0$ should be generally suppressed relative to $\beta =0$ and replaced by dynamical localization for irrational $\beta$. This can be clearly seen in Fig. 2. The inset of this figure shows that the quantum-resonance suppression is visible for experimentally realistic values \cite{e2} of quasimomentum uncertainty $\Delta\beta$ and number of kicks.

\begin{figure}[tbp]
\includegraphics[width=7.5cm]{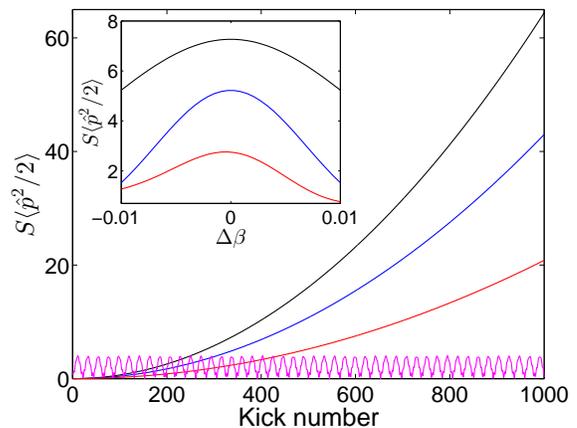}
\caption{(Color online) Scaled mean kinetic energy $S\langle \hat{p}^2/2\rangle$ ($S$ is a scaling factor) vs. kick number for a zero-momentum initial state, $k=5$, $V(x)=\cos (x)$, $\hbar _{\mathrm{s}}=1/2$, and, in order of descending curves: $\beta =0$, $S=4\cdot 10^{-4}$; $\beta =1/2$, $S=4\cdot 10^{-3}$; $\beta =1/3$, $S=0.01$; $\beta =2-\sqrt{3}$, $S=1$ (dynamical localization). The inset shows $S\langle \hat{p}^2/2\rangle$ after $30$ kicks vs. $\Delta\beta =\beta-\beta_0$ in the first three cases above: $\beta_0 =0$, $S=0.05$; $\beta_0 =1/2$, $S=0.5$; $\beta_0 =1/3$, $S=0.75$ (in order of descending curves).}
\label{fig2}
\end{figure}
The dynamical localization for irrational $\beta$ is basically different from the usual one for irrational $\hbar _{\mathrm{s}}$ \cite{dl,al} in several aspects. First, all the QE eigenstates (\ref{pprl}) are obtained by applying the momentum translations $\exp (irqx)$ [with $n\hbar\rightarrow (n-rq)\hbar$] to just $q$ independent eigenstates corresponding to the $q$ independent values of $\omega$ giving the ladders (\ref{wr}). This implies, e.g., that two initial momentum states differing by $r'q\hbar$ ($r'$ integer) will evolve to states whose localized momentum probability distributions coincide up to a momentum shift of $r'q\hbar$. This is illustrated in Fig. 3 for $q=3$ and $r'=5$, assuming a realistic quasimomentum width $\Delta\beta$ for BECs.

\begin{figure}[tbp]
\includegraphics[width=7.5cm]{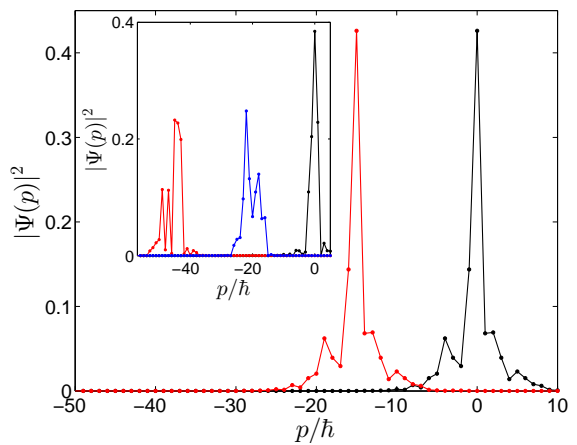}
\caption{(Color online) Dynamically-localized momentum probability distributions, averaged over a quasimomentum width of $\Delta\beta =0.06$, evolved after $20$ kicks from the two momentum states with $p=0$ and $p=-15\hbar$ for $k=5$, $V(x)=\cos (x)$, $\hbar _{\mathrm{s}}=1/3$, and $\beta =2-\sqrt{3}$. Clearly, the distributions coincide up to a momentum shift of $15\hbar$. The inset shows the localized momentum probability distributions of the $q=3$ independent QE eigenstates in this case.}
\label{fig3}
\end{figure}      
Second, the expectation value of any physical observable for irrational $\beta$ will evolve in time with a frequency (Fourier) spectrum $\nu$ given by the differences of all $q$ ladders (\ref{wr}): $\nu =\omega_c-\omega_{c'}+2\pi rl(\beta +q/2)$ mod($2\pi$); here $\omega_c$, $\omega_{c'}$ are any two of the $q$ independent levels $\omega$ defining the ladders (\ref{wr}) and $r$ is an arbitrary integer. This gives a staggered-ladder frequency spectrum which is symmetric around the central ladder for $c=c'$: $\nu=2\pi rl(\beta +q/2)$ mod($2\pi$). Clearly, this ladder is completely independent of the nonintegrability, as illustrated in Fig. 4.

\begin{figure}[tbp]
\includegraphics[width=7.5cm]{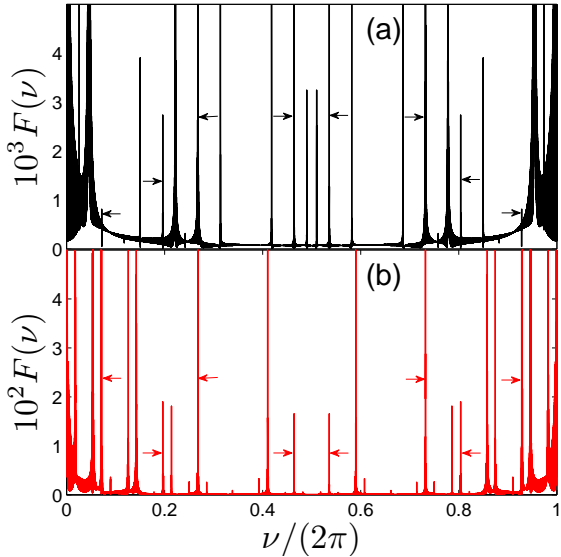}
\caption{(Color online) Fourier transform $F(\nu )$ of the mean kinetic energy $\langle \hat{p}^2/2\rangle$ vs. kick number for a zero-momentum initial state, $V(x)=\cos (x)$, $\hbar _{\mathrm{s}}=1/2$, $\beta =2-\sqrt{3}$, and: (a) $k=5$; (b) $k=8$. The arrows indicate the central ladder of frequencies $\nu =2\pi r\beta$ mod($2\pi$), $r=\pm 1,...,\pm 4$, clearly independent of the nonintegrability $k$.}
\label{fig4}
\end{figure}
Third, consider the most general wavepacket associated with a ladder subspectrum (\ref{wr}). This is an arbitrary linear combination of the eigenstates (\ref{pprl}): 
\begin{equation}
\phi _{\beta}(x)=\psi _{\beta ,\omega }(x)\chi (x),
\label{aiwp}
\end{equation}
where
\begin{equation}
\chi (x) = \sum_{r=-\infty }^{\infty }c_{r}\exp (irqx) \label{chi}
\end{equation} 
and $c_r$ are arbitrary coefficients. Using Eq. (\ref{wr}) and the fact that (\ref{pprl}) is an eigenstate of $\hat{U}_{\beta}$ with eigenvalue $\exp (-i\omega_{rl})$, we find that after $s$ kicks the wavepacket (\ref{aiwp}) with (\ref{chi}) will evolve to
\begin{equation}
\phi _{\beta}(x;s)=\hat{U}_{\beta}^{s}\phi _{\beta}(x) = e^{-i\omega s}\psi _{\beta ,\omega }(x)\chi (x-\beta ^{\prime }\hbar s),  \label{awp}
\end{equation}   
where $\beta ^{\prime }=\beta +q/2\ \mathrm{mod}(1)$. Thus, the evolving wavepacket (\ref{awp}) contains a traveling-wave component $\chi (x-\beta ^{\prime }\hbar s)$ moving without change of shape at constant velocity $\beta ^{\prime}\hbar$. We now show that this component is clearly exhibited by $\vert\phi _{\beta }(x;s)\vert ^2$, without being masked by $\psi _{\beta ,\omega }(x)$ in Eq. (\ref{awp}), in at least two cases. The first case is that of the main quantum resonances, $\hbar _{\mathrm{s}}=l$ ($q=1$). Since the eigenvalues of $\hat{p}/\hbar$ in Eq. (\ref{ub}) are integers, it is easy to see that $\hat{U}_\beta$ can be expressed in this case as
\begin{equation}
\hat{U}_\beta=\exp (-i\pi l\beta^2)\exp (-i\beta ^{\prime }\hat{p})\exp [-ikV(\hat{x})/\hbar]. \label{ub1}
\end{equation}
The second exponential in Eq. (\ref{ub1}) is just a translation in $x$ by $-\beta ^{\prime }\hbar$. We then get the exact result
\begin{equation}
\phi _{\beta}(x;s)=\hat{U}_{\beta}^{s}\phi _{\beta}(x) = \exp [-i\eta_{\beta}(x;s)]\phi _{\beta}(x-\beta ^{\prime }\hbar s),  \label{awp1}
\end{equation}  
where $\eta_{\beta}(x;s)=\pi ls\beta^2+k\sum_{m=1}^sV(x-\beta ^{\prime }\hbar m)/\hbar$. Thus, one always has the exact traveling-wave behavior $\vert\phi _{\beta }(x;s)\vert ^2=\vert\phi _{\beta}(x-\beta ^{\prime }\hbar s)\vert ^2$ for general initial wavepacket $\phi _{\beta}(x)=\chi(x)$, given by Eq. (\ref{chi}) with $q=1$.
 
The second case is that of arbitrary $q$ for sufficiently small nonintegrability $k$. In this case, the dynamical-localization length for irrational $\beta$ will be small and the QE eigenstate $\psi _{\beta ,\omega }(x)$ in Eqs. (\ref{aiwp}) and (\ref{awp}) will be close to a pure angular-momentum state $\exp (inx)$. Then, starting from the state $\phi_{\beta}(x)=\exp (inx)\chi (x)$, for any integer $n$ and arbitrary $\chi (x)$ in Eq. (\ref{chi}), $\vert\phi_{\beta}(x;s)\vert ^2$ should exhibit approximately a traveling-wave evolution. Fig. 5 shows that this evolution indeed holds to a good approximation in the two cases considered even when a realistic quasimomentum width of BECs is assumed. The simple initial states used in Fig. 5 are superpositions of two momentum states and can be experimentally realized as described in works \cite{e10,e3}.

\begin{figure}[tbp]
\includegraphics[width=7.5cm]{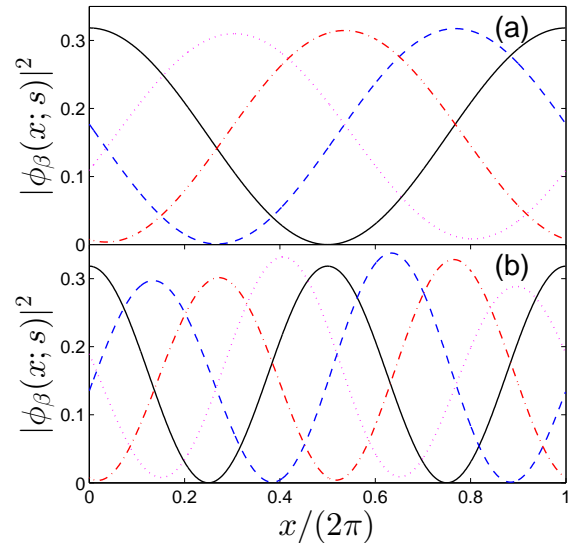}
\caption{(Color online) (a) Time ($s$) evolution of $|\phi_{\beta}(x;s)|^{2}$,  averaged over a quasimomentum width of $\Delta\beta =0.06$, for $k=10$, $V(x)=\cos (x)$, $\hbar _{\mathrm{s}}=1$, $\beta =2-\sqrt{3}$, initial state $[1+\exp (-x)]/\sqrt{4\pi}$, and: $s=0$ (solid line), $s=1$ (blue dashed line), $s=2$ (red dot-dashed line), $s=3$ (magenta dotted line). (b) Similar to (a) but with the following changes: $k=0.1$, $\hbar _{\mathrm{s}}=1/2$ ($q=2$), and initial state $[1+\exp (-2x)]/\sqrt{4\pi}$. In both cases, especially in case (a) for $q=1$, we see that a traveling-wave evolution occurs to a good approximation.}
\label{fig5}
\end{figure}
In summary, we have presented the first case of a staggered-ladder QE spectrum, exhibited by the paradigm (\ref{KP}) of classically nonintegrable systems under the experimentally realizable conditions of quantum resonance for generic quasimomentum $\beta$; the case of $\beta =0$ (usual kicked rotor) turns out to be non-generic. The new regular QE structure implies novel quantum-dynamical phenomena which persist when averaged over realistic quasimomentum widths $\Delta\beta$ of BECs and should therefore be experimentally observable.     
 
We thank S. Fishman, G.S. Summy, and M. Wilkinson for discussions. This work was partially supported by Bar-Ilan University Grant No. 2046.

\end{document}